\documentclass[prd,twocolumn,superscriptaddress,showpacs,amssymb,amsmath,amsfonts,aps,altaffilletter]{revtex4}

\usepackage{color}
\usepackage{times}
\usepackage{graphicx}
\usepackage{fancyhdr}
\usepackage{float}
\usepackage{ulem}
\usepackage[raggedright]{subfigure}
\usepackage{acronym}
\usepackage{multirow}
\usepackage{array}

\makeatletter
\makeatletter
\def\@fnsymbol#1{\ifcase#1\or * \or  $+$ \or  \$ \or \#  \or \dag \or \ddag \or
$\mathsection$ \or $ \mathparagraph$ \or $\|$  \or \textordfeminine \or \textbul
let   
\or ** \or $++$ \or  \$\$ \or \#\#  \or \dag\dag \or \ddag\ddag \or
$\mathsection\mathsection$ \or $ \mathparagraph\mathparagraph$ \or $\|\|$  \or 
\textordfeminine\textordfeminine \or \textbullet \textbullet \or *** \or $+++$ 
\or  \$\$\$ \or \#\#  \or \dag\dag \or \ddag\ddag \or
$\mathsection \mathsection\mathsection$ \or $ \mathparagraph 
\mathparagraph\mathparagraph$ \or $\|\|\|$  \or 
\textordfeminine\textordfeminine\textordfeminine \or 
\textbullet\textbullet\textbullet \or \else \@ctrerr\fi}
\makeatother

\newcommand\fake[1]{\textcolor{red}{#1}}

\def\thercsid{\relax}
\def\rcsid#1{\def\next##1#1{\def\thercsid{##1}}\next}

\rcsid$Id: s5-1yr-lowcbc.tex,v 1.103 2009/04/28 20:29:00 dkeppel Exp $

\renewcommand{\today}{\number\day\space\ifcase\month\or
  January\or February\or March\or April\or May\or June\or
  July\or August\or September\or October\or November\or December\fi
  \space\number\year}

\def\Msun{\ensuremath{M_{\odot}}}

\def\BNSul{\ensuremath{3.9 \times 10^{-2}}}
\def\SBHNSul{\ensuremath{1.4 \times 10^{-2}}}
\def\BHNSul{\ensuremath{1.1 \times 10^{-2}}}
\def\SBBHul{\ensuremath{3.2 \times 10^{-3}}}
\def\BBHul{\ensuremath{2.5 \times 10^{-3}}}

\begin{document}

\acrodef{BBH}{binary black holes}
\acrodef{BNS}{binary neutron stars}
\acrodef{BHNS}{black hole-neutron star binaries}
\acrodef{PBH}{primordial black hole binaries}
\acrodef{SNR}{signal-to-noise ratio}
\acrodef{SPA}{stationary-phase approximation}
\acrodef{LIGO}{Laser Interferometer Gravitational-wave Observatory}
\acrodef{LHO}{LIGO Hanford Observatory}
\acrodef{LLO}{LIGO Livingston Observatory}
\acrodef{LSC}{LIGO Scientific Collaboration}
\acrodef{GRB}{gamma-ray bursts}
\acrodef{CBC}{compact binary coalescences}
\acrodef{GW}{gravitational waves}
\acrodef{ISCO}{innermost stable circular orbit}
\acrodef{FAR}{false alarm rate}
\acrodef{IFAR}{inverse false alarm rate}
\acrodef{CL}{confidence level}
\acrodef{PN}{post-Newtonian}
\acrodef{DQ}{data quality}

\title{Search for Gravitational Waves from Low Mass Binary Coalescences in the
First Year of LIGO's S5 Data\\
}

%
%
%
%

\newcommand*{\AG}{Albert-Einstein-Institut, Max-Planck-Institut f\"{u}r Gravitationsphysik, D-14476 Golm, Germany}
\affiliation{\AG}
\newcommand*{\AH}{Albert-Einstein-Institut, Max-Planck-Institut f\"{u}r Gravitationsphysik, D-30167 Hannover, Germany}
\affiliation{\AH}
\newcommand*{\AU}{Andrews University, Berrien Springs, MI 49104 USA}
\affiliation{\AU}
\newcommand*{\AN}{Australian National University, Canberra, 0200, Australia}
\affiliation{\AN}
\newcommand*{\CH}{California Institute of Technology, Pasadena, CA  91125, USA}
\affiliation{\CH}
\newcommand*{\CA}{Caltech-CaRT, Pasadena, CA  91125, USA}
\affiliation{\CA}
\newcommand*{\CU}{Cardiff University, Cardiff, CF24 3AA, United Kingdom}
\affiliation{\CU}
\newcommand*{\CL}{Carleton College, Northfield, MN  55057, USA}
\affiliation{\CL}
\newcommand*{\CS}{Charles Sturt University, Wagga Wagga, NSW 2678, Australia}
\affiliation{\CS}
\newcommand*{\CO}{Columbia University, New York, NY  10027, USA}
\affiliation{\CO}
\newcommand*{\ER}{Embry-Riddle Aeronautical University, Prescott, AZ   86301 USA}
\affiliation{\ER}
\newcommand*{\EU}{E\"{o}tv\"{o}s University, ELTE 1053 Budapest, Hungary}
\affiliation{\EU}
\newcommand*{\HC}{Hobart and William Smith Colleges, Geneva, NY  14456, USA}
\affiliation{\HC}
\newcommand*{\IA}{Institute of Applied Physics, Nizhny Novgorod, 603950, Russia}
\affiliation{\IA}
\newcommand*{\IU}{Inter-University Centre for Astronomy  and Astrophysics, Pune - 411007, India}
\affiliation{\IU}
\newcommand*{\HU}{Leibniz Universit\"{a}t Hannover, D-30167 Hannover, Germany}
\affiliation{\HU}
\newcommand*{\CT}{LIGO - California Institute of Technology, Pasadena, CA  91125, USA}
\affiliation{\CT}
\newcommand*{\LO}{LIGO - Hanford Observatory, Richland, WA  99352, USA}
\affiliation{\LO}
\newcommand*{\LV}{LIGO - Livingston Observatory, Livingston, LA  70754, USA}
\affiliation{\LV}
\newcommand*{\LM}{LIGO - Massachusetts Institute of Technology, Cambridge, MA 02139, USA}
\affiliation{\LM}
\newcommand*{\LU}{Louisiana State University, Baton Rouge, LA  70803, USA}
\affiliation{\LU}
\newcommand*{\LE}{Louisiana Tech University, Ruston, LA  71272, USA}
\affiliation{\LE}
\newcommand*{\LL}{Loyola University, New Orleans, LA 70118, USA}
\affiliation{\LL}
\newcommand*{\MT}{Montana State University, Bozeman, MT 59717, USA}
\affiliation{\MT}
\newcommand*{\MS}{Moscow State University, Moscow, 119992, Russia}
\affiliation{\MS}
\newcommand*{\ND}{NASA/Goddard Space Flight Center, Greenbelt, MD  20771, USA}
\affiliation{\ND}
\newcommand*{\NA}{National Astronomical Observatory of Japan, Tokyo  181-8588, Japan}
\affiliation{\NA}
\newcommand*{\NO}{Northwestern University, Evanston, IL  60208, USA}
\affiliation{\NO}
\newcommand*{\RI}{Rochester Institute of Technology, Rochester, NY  14623, USA}
\affiliation{\RI}
\newcommand*{\RA}{Rutherford Appleton Laboratory, HSIC, Chilton, Didcot, Oxon OX11 0QX United Kingdom}
\affiliation{\RA}
\newcommand*{\SJ}{San Jose State University, San Jose, CA 95192, USA}
\affiliation{\SJ}
\newcommand*{\SM}{Sonoma State University, Rohnert Park, CA 94928, USA}
\affiliation{\SM}
\newcommand*{\SE}{Southeastern Louisiana University, Hammond, LA  70402, USA}
\affiliation{\SE}
\newcommand*{\SO}{Southern University and A\&M College, Baton Rouge, LA  70813, USA}
\affiliation{\SO}
\newcommand*{\SA}{Stanford University, Stanford, CA  94305, USA}
\affiliation{\SA}
\newcommand*{\SR}{Syracuse University, Syracuse, NY  13244, USA}
\affiliation{\SR}
\newcommand*{\PU}{The Pennsylvania State University, University Park, PA  16802, USA}
\affiliation{\PU}
\newcommand*{\UM}{The University of Melbourne, Parkville VIC 3010, Australia}
\affiliation{\UM}
\newcommand*{\MI}{The University of Mississippi, University, MS 38677, USA}
\affiliation{\MI}
\newcommand*{\SF}{The University of Sheffield, Sheffield S10 2TN, United Kingdom}
\affiliation{\SF}
\newcommand*{\TA}{The University of Texas at Austin, Austin, TX 78712, USA}
\affiliation{\TA}
\newcommand*{\TC}{The University of Texas at Brownsville and Texas Southmost College, Brownsville, TX  78520, USA}
\affiliation{\TC}
\newcommand*{\TR}{Trinity University, San Antonio, TX  78212, USA}
\affiliation{\TR}
\newcommand*{\BB}{Universitat de les Illes Balears, E-07122 Palma de Mallorca, Spain}
\affiliation{\BB}
\newcommand*{\UA}{University of Adelaide, Adelaide, SA 5005, Australia}
\affiliation{\UA}
\newcommand*{\BR}{University of Birmingham, Birmingham, B15 2TT, United Kingdom}
\affiliation{\BR}
\newcommand*{\FA}{University of Florida, Gainesville, FL  32611, USA}
\affiliation{\FA}
\newcommand*{\GU}{University of Glasgow, Glasgow, G12 8QQ, United Kingdom}
\affiliation{\GU}
\newcommand*{\MD}{University of Maryland, College Park, MD 20742 USA}
\affiliation{\MD}
\newcommand*{\AM}{University of Massachusetts - Amherst, Amherst, MA 01003, USA}
\affiliation{\AM}
\newcommand*{\MU}{University of Michigan, Ann Arbor, MI  48109, USA}
\affiliation{\MU}
\newcommand*{\MN}{University of Minnesota, Minneapolis, MN 55455, USA}
\affiliation{\MN}
\newcommand*{\OU}{University of Oregon, Eugene, OR  97403, USA}
\affiliation{\OU}
\newcommand*{\RO}{University of Rochester, Rochester, NY  14627, USA}
\affiliation{\RO}
\newcommand*{\SL}{University of Salerno, 84084 Fisciano (Salerno), Italy}
\affiliation{\SL}
\newcommand*{\SN}{University of Sannio at Benevento, I-82100 Benevento, Italy}
\affiliation{\SN}
\newcommand*{\SH}{University of Southampton, Southampton, SO17 1BJ, United Kingdom}
\affiliation{\SH}
\newcommand*{\SC}{University of Strathclyde, Glasgow, G1 1XQ, United Kingdom}
\affiliation{\SC}
\newcommand*{\WA}{University of Western Australia, Crawley, WA 6009, Australia}
\affiliation{\WA}
\newcommand*{\UW}{University of Wisconsin-Milwaukee, Milwaukee, WI  53201, USA}
\affiliation{\UW}
\newcommand*{\WU}{Washington State University, Pullman, WA 99164, USA}
\affiliation{\WU}

\author{}    \affiliation{\GU}    
\author{B.~P.~Abbott}    \affiliation{\CT}    
\author{R.~Abbott}    \affiliation{\CT}    
\author{R.~Adhikari}    \affiliation{\CT}    
\author{P.~Ajith}    \affiliation{\AH}    
\author{B.~Allen}    \affiliation{\AH}  \affiliation{\UW}  
\author{G.~Allen}    \affiliation{\SA}    
\author{R.~S.~Amin}    \affiliation{\LU}    
\author{S.~B.~Anderson}    \affiliation{\CT}    
\author{W.~G.~Anderson}    \affiliation{\UW}    
% ...
 \collaboration{The LIGO Scientific Collaboration, http://www.ligo.org}
 \noaffiliation
%
%

\fake{\pacs{95.85.Sz, 04.80.Nn, 07.05.Kf, 97.60.Jd, 97.60.Lf, 97.80.-d}}

\begin{abstract}\quad
We have searched for gravitational waves from coalescing
low mass compact binary systems with a total mass between $2$ and $35~\Msun$
and a minimum component mass of $1~\Msun$ using data from the first year of the
fifth science run (S5) of the three LIGO detectors,
operating at design sensitivity.
Depending on mass, we are sensitive to coalescences
as far as 150 Mpc from the Earth.
No gravitational wave signals were observed above the expected background.
Assuming a compact binary objects population with a Gaussian mass distribution
representing binary neutron star systems,
black hole-neutron star binary systems,
and binary black hole systems,
we calculate the 90\%-confidence upper limit on the rate of coalescences
to be \BNSul $\textrm{ yr}^{-1} L_{10}^{-1}$,
\BHNSul $\textrm{ yr}^{-1} L_{10}^{-1}$,
and \BBHul $\textrm{ yr}^{-1} L_{10}^{-1}$ respectively,
where $L_{10}$ is $10^{10}$ times the blue solar luminosity.
We also set improved upper limits on the rate of 
compact binary coalescences per unit blue-light luminosity,
as a function of mass.
\end{abstract}

\maketitle

\section{Introduction}\label{sec:overview}

Among the most promising candidates for the first detection of \ac{GW} 
are signals from \ac{CBC}, 
which include \ac{BNS}, \ac{BBH}, and \ac{BHNS}. 
The inspiral waveforms generated by these systems can be 
reliably predicted using \ac{PN} perturbation theory, 
until the last fraction of a second prior to merger.
These waveforms can be used in matched filtering of 
noisy data from gravitational wave detectors
to identify \ac{GW} candidate events.

Astrophysical estimates for \ac{CBC} rates depend on a number of assumptions
and unknown model parameters, and are still uncertain at present.
In the simplest models, the coalescence rates should be
proportional to the stellar birth rate in nearby spiral galaxies,
which can be estimated from their blue luminosity \cite{LIGOS3S4Galaxies};
we therefore express the coalescence rates per unit $L_{10}$,
where $L_{10}$ is $10^{10}$ times the blue solar luminosity
(the Milky Way contains $\sim1.7 L_{10}$ \cite{Kalogera:2000dz}).
The most confident \ac{BNS} rate predictions are based on extrapolations
from observed binary pulsars in our Galaxy;
these yield realistic \ac{BNS} rates of $5 \times 10^{-5} \textrm{ yr}^{-1} L_{10}^{-1}$,
although rates could plausibly be as high as
$5 \times 10^{-4} \textrm{ yr}^{-1} L_{10}^{-1}$ \cite{Kalogera:2004tn, Kalogera:2004nt}.
Predictions for \ac{BBH} and \ac{BHNS} rates are based on population synthesis
models constrained by these and other observations.
Realistic rate estimates are
$2 \times 10^{-6} \textrm{ yr}^{-1} L_{10}^{-1}$ for \ac{BHNS} \cite{Oshaughnessy:2008} and
$4 \times 10^{-7} \textrm{ yr}^{-1} L_{10}^{-1}$ for \ac{BBH} \cite{OShaughnessy:2005};
both \ac{BHNS} and \ac{BBH} rates could plausibly be as high as
$6 \times 10^{-5} \textrm{ yr}^{-1} L_{10}^{-1}$
\cite{Oshaughnessy:2008, OShaughnessy:2005}.

The \ac{LIGO} detectors achieved design sensitivity
in 2005, and completed a two-year-long science run
(S5) in November 2007. 
Results from searches for \ac{GW} from \ac{CBC}
by the \ac{LSC} using data
from previous science runs with ever-increasing sensitivity
are reported in Refs.~\cite{LIGOS1iul,LIGOS2iul,LIGOS2macho,LIGOS2bbh,LIGOS3S4all}.

This paper summarizes the search 
for \ac{GW} signals from \ac{CBC} with 
component masses greater than or equal to 1 solar mass (\Msun)
and total mass ranging from $2$ to $35~\Msun$,
using the first year of data from \ac{LIGO}'s S5 run,
between November 4th, 2005 and November 14th, 2006.
During this time, the \ac{LIGO} detectors 
were sensitive to signals from \ac{CBC} with horizon distances
(Table \ref{tab:ng})
of 30 Mpc for \ac{BNS}
(25 seconds in the LIGO detection band)
and 150 Mpc for 
systems with a total mass of $\sim 28~\Msun$
(0.5 seconds in the LIGO band).
Subsequent papers will report the results of similar searches
using the data from the second year
(during which time the Virgo detector was in observational mode),
searches for higher-mass systems (between $25$ and $100~\Msun$),
and specialized searches targeting particular subsets of signals.

The component objects of true astrophysical compact binaries 
will in general have some angular momentum,
for which \ac{PN} waveforms that incorporate non-zero
values for the spin parameters are available 
\cite{BuonannoChenVallisneri:2003b,S3_BCVSpin}.
However, for most of the parameter space, the effect of spin on the waveforms
is small, and the signals can be captured using non-spinning
waveform templates (Appendix \ref{appendix:BCVSpin})
with only a small loss in the \ac{SNR};
this is the approach taken in the search described here.
In some regions of parameter space, the effect of spin is larger,
and dedicated searches 
\cite{S3_BCVSpin,Pan:2003qt,Buonanno:2004yd,Buonanno:2005pt} 
may be more effective.
The \ac{LSC} continues to develop more effective methods
for searching for signals with strong modulations due to spin.

The rest of this paper is organized as follows. Section \ref{sec:pipeline}
summarizes the search pipeline that was employed.
Section \ref{sec:detection} describes the output of the search:
detection candidate events which are examined and rejected
using a detection confidence procedure.
Section \ref{sec:efficiency} describes the evaluation of our
detection efficiency using simulated \ac{GW} signals
injected into the detectors' data streams.
Section \ref{sec:ul} discusses the upper limit calculation that was performed,
and the resulting upper limits on \ac{CBC} rates neglecting spin of the
coalescing objects.
Section \ref{sec:spininj} discusses how our sensitivity is affected
when spin is included.
Finally, Section \ref{sec:conc} presents the conclusions, followed by
several appendices on certain technical aspects of the search.

\section{The Data Analysis Pipeline}\label{sec:pipeline}

The pipeline used for this analysis has been described in previous
documents \cite{findchirppaper,LIGOS3S4Tuning,systematics,BBCCS:2006},
and was used to search for \ac{BNS} in \ac{LIGO}'s 
third and fourth science runs \cite{LIGOS3S4all}.
The main aspects of the pipeline and new features used here are detailed below.

The data analysis proceeds as follows.
The gravitational wave strain data
are recorded from each of the three LIGO detectors:
the H1 and H2 detectors at \ac{LHO} and the L1 detector at \ac{LLO}.
These data are matched filtered through banks of 
templates that model the expected signal
from a binary coalescence
of two compact massive objects with masses $m_1$ and $m_2$,
resulting in triggers that pass a pre-set \ac{SNR} threshold.
We search for coincident triggers in time and template masses,
between two or three detectors.
We subject these coincident triggers
to several tests to suppress noise fluctuations
(including the $\chi^2$ test described in \cite{Allen:2004}),
and rank-order the remaining coincident triggers according to
their inconsistency with the background.

We estimate the background from accidental coincidences
by looking at time-shifted coincident triggers, as detailed in
Section \ref{subsec:background}.
Coincident triggers that are not consistent with the
estimated background are followed up with many additional
consistency checks, designed to identify strong but rare
noise fluctuations.
We estimate our sensitivity to \ac{GW} signals
through injections of simulated waveforms into the 
\ac{LIGO} data stream which are analyzed identically to the data.

\subsection{Template Bank}\label{subsec:tmplts}
The templates used for this search are waveforms from non-spinning compact
binaries 
calculated in the frequency domain using the 
\ac{SPA} \cite{thorne.k:1987,SathyaDhurandhar:1991,Droz:1999qx}.
The waveforms are calculated to
Newtonian order in amplitude and second \ac{PN} order in
phase, and extend until the Schwarzschild \ac{ISCO}.
The templates for this single search cover a larger binary mass region
than in previous searches \cite{LIGOS3S4all}, with a total mass ($M$) of 
$2 M_{\odot}< M < 35 M_{\odot}$ and a minimum component mass of 
$1~\Msun$.  The templates are placed with a hexagonal spacing \cite{hexabank}
such that we lose less than $3\%$ of the \ac{SNR} due to
using a discrete template bank to cover the continuous parameter space
spanned by the two component masses.

\subsection{Analyzed and Vetoed Times}
The pipeline is applied to data from the first year of the
LIGO S5 run for which more than one detector was in observation mode.
This comprises 0.419 yr of triple coincident data (H1H2L1),
0.232 yr of H1H2 coincident data, 0.037 yr of H1L1 coincident data,
and 0.047 yr of H2L1 coincident data.
In determining our upper limits,
we exclude approximately 9.5\%\ of the data that were used
to tune the pipeline \cite{LIGOS3S4Tuning} (the \textit{playground} data).
We also exclude all the data when only the H1 and H2 detectors
were in observation mode, because of the difficulty in
determining the background from coincident noise triggers
in these collocated detectors (Section \ref{subsec:background}).
We make use of the (rather large amount of) additional information
on the state of the detectors and the physical environment
to define \ac{DQ} criteria (Appendix \ref{appendix:dataqual}).  
We use these \ac{DQ} criteria to veto triggers in times when an individual 
detector was in observation mode, but we have reason to believe the data were 
contaminated by instrumental or environmental problems.  
We define four categories of vetoes from these \ac{DQ} criteria,
based on severity of the data quality issue,
and how well we understand its origin,
explained in Appendix \ref{appendix:dataqual}.
We follow up detection candidates after successively applying each
veto category (Appendix \ref{appendix:detection_checklist}).
We exclude from the upper limit calculation times flagged with \ac{DQ} vetoes
in the first three categories, along with triggers recorded in those times.
This results in non-vetoed, non-playground observation times of
0.336 yr for H1H2L1, 0.020 yr for H1L1, and 0.041 yr for H2L1.

\subsection{Coincidence Test and Clustering}\label{subsec:coincidence}
Our analysis applies a more sophisticated coincidence test than the one used
in the past.
Previously, in order for triggers from different interferometers to be
considered \textit{coincident triggers},
they needed to pass a series of independent windows in time, chirp mass
($M_{\rm c} = \eta^{3/5} M$),
and symmetric mass ratio
($\eta = m_1 m_2 / M^2$).
These windows were defined independently of the parameters of the triggers
(e.g.~$M_{\rm c}$, $\eta$).

We have replaced this coincidence test with the one \cite{Robinson:2008}
that is based on the metric used in constructing a template
bank \cite{Balasubramanian:1995bm,Owen:1995tm,Owen:1998dk,BBCCS:2006}.
The metric contains terms necessary for measuring distances and
determining coincidence in masses
and time as well as the correlations between the parameters
expected for real signal events in the 3-dimensional parameter space.
This provides improved separation between signals
and background from accidental coincidence of noise triggers,
compared to the above independent windows.

We have also changed the algorithm used to cluster single-detector
triggers in our pipeline.
Previously, the triggers were clustered by retaining the trigger with the
largest \ac{SNR} from all the templates over a fixed window of time.
At present, we use a new method \cite{SenguptaTrigScan:2008}
to cluster triggers, analogous to the coincidence algorithm,
again retaining the trigger with the largest \ac{SNR}
from a particular cluster.

\subsection{Background Estimation}\label{subsec:background}
As in the previous searches,
we estimate the background due to accidental coincidences of noise triggers
by repeating the analysis with the triggers from
different detectors shifted in time relative to each other,
forming 100 experimental trials with no true signals.
We refer to these as \textit{time-shifted coincident triggers},
as opposed to the \textit{in-time coincident triggers}
obtained without the use of time-shifts.

This procedure is known to underestimate the rate of 
accidental coincidences of noise triggers from the
H1 and H2 detectors, since they are collocated and 
exhibit time-correlated noise excursions.
We therefore exclude H1H2 double-coincident data from the
upper limit calculation.
We examine only the very strongest H1H2 double-coincident
detection candidates (including H1H2 coincidences that did not appear in L1),
and subject them to very stringent scrutiny.
There were no H1H2 candidates that survived these checks.
(See Section \ref{sec:detection} for details.)

\subsection{Detection Statistic}\label{subsec:detstat}
In this search, we employ a new detection statistic which allows
us to search over a large region of parameter space without
being limited by a high background \ac{FAR} from a smaller subregion.
In Ref.~\cite{LIGOS3S4all}, coincident triggers were ranked by 
combined effective \ac{SNR} (Appendix \ref{appendix:effsnr}).
Here, instead, we use a statistic derived from the background
\ac{FAR}, as detailed in Appendix \ref{appendix:FAR}.
The time-shifted triggers provide an estimate
of the \ac{FAR} for each in-time coincident trigger.
By counting the number of time-shifted triggers with an effective \ac{SNR}
greater than or equal to the in-time coincident triggers' effective \ac{SNR},
and dividing by the total amount of time we searched for time-shifted triggers,
we calculate the \ac{FAR} for each in-time coincident trigger.
This procedure is done separately for different regions of parameter
space with the result that the \ac{FAR} as a function of effective \ac{SNR}
varies over the parameter space.
In-time coincident triggers with the largest \ac{IFAR} 
are our best detection candidates.

\section{Detection candidates}\label{sec:detection}

\begin{figure*}[t]
\includegraphics[width=0.9\textwidth]{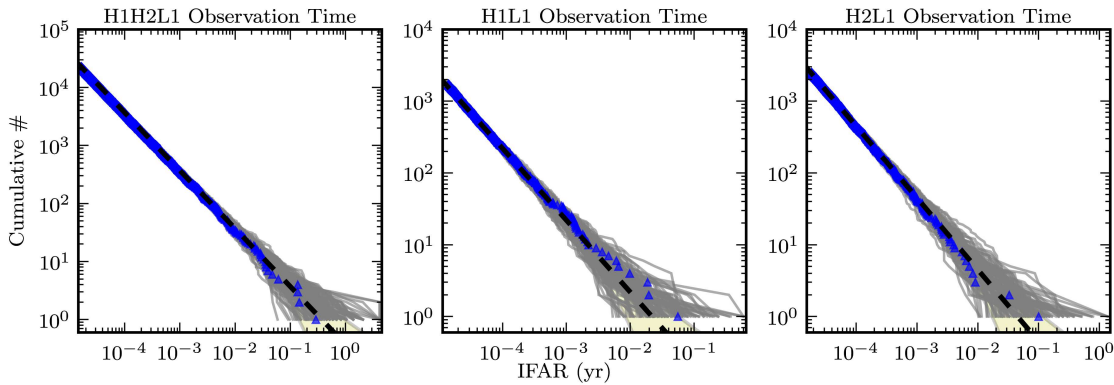}
\caption{
The cumulative distribution of events above a threshold \ac{IFAR},
for in-time coincident events, shown as blue triangles,
from all coincidence categories for the observation times
H1H2L1, H1L1, and H2L1 respectively.
The expected background (by definition) is shown as a dashed black line.
The 100 experimental trials that make up our background are also
plotted individually as the solid grey lines.
The shaded region denotes the $N^{1/2}$ errors.
}
\label{fig:IFAR}
\end{figure*}

At the end of our pipeline we are left with a set of coincident triggers
that are potential detection candidates.
The cumulative distribution of events above a threshold \ac{IFAR}
is shown in Fig.~\ref{fig:IFAR}.
This figure shows that the loudest candidates in all three sets were
consistent with the estimated background and thus were likely accidental
coincidences.
Thus, the search yielded no detection candidates,
and we report an upper limit in Sections \ref{sec:ul} and \ref{sec:spininj}.

As an exercise to prepare for future detections,
we carry the loudest several events (such as the three loudest events
that appear in each of the histograms in Fig.~\ref{fig:IFAR})
through a detection checklist described in
Appendix \ref{appendix:detection_checklist}.
The methods employed in this checklist are tested against
simulated \ac{GW} signals and the time-shifted coincidence triggers
used to estimate the background.

Even though we know our background is underestimated for H1H2 coincident
triggers, we reviewed the two loudest H1H2 candidates using the
detection checklist.
In both of those cases, the waveforms from the two interferometers
failed to match each other in detail,
thus ruling them out as gravitational wave events.

During the analysis, and prior to unblinding the non-playground data, an error
was found in the coincidence algorithm (Appendix
\ref{appendix:detection_checklist}). This caused the coincidence requirement
to be tighter than initially intended.  It had a negligible effect for low
mass templates, but became more significant at higher masses.  However, since
the coincidence threshold was selected based upon the examination of simulated
signals, we decided to use this search in generating the upper limits
presented here.  We verified the detection candidates by re-running the search
after correcting the coincidence test.  The results of the corrected search
did not provide any plausible gravitational-wave signals.

\section{Detection efficiency}\label{sec:efficiency}

We evaluate our efficiency for detecting \ac{GW} signals
from \ac{CBC} during the first year of S5 as a function of 
mass and of distance to the source.
This is done by coherently injecting a large number of simulated signals,
called software injections,
into the detector data streams.
Those data are then analyzed with a pipeline identical 
to that used to search for detection candidates.
The distribution of masses, distances, 
sky locations, orientations, and component spins
is described in Appendix \ref{appendix:injections}.
The procedure for calculating the detection efficiency
is described in Appendix \ref{appendix:chirp_distance},
where a software injection is considered to be detected if its \ac{IFAR}
exceeds that of the coincident in-time trigger with
the highest \ac{IFAR}.
We find that our detection efficiencies are consistent
with expectations from the detectors' noise spectra during S5.

As noted in Sections \ref{sec:overview} 
and \ref{sec:pipeline},
we are using non-spinning templates to look for
\ac{GW} from \ac{CBC}, whereas 
true \ac{GW} signals from \ac{CBC} will have some amount
of spin associated with the objects. 
Therefore, in the next two sections,
we evaluate our detection efficiency using injections of both non-spinning
and spinning simulated signals.

Appendix \ref{appendix:BCVSpin} describes a comparison of the
pipeline described above with one using 
phenomenological waveforms \cite{S3_BCVSpin,Pan:2003qt,Buonanno:2004yd,Buonanno:2005pt, VDBroeck:2008}.
The present pipeline admits the use of the $\chi^2$ test \cite{Allen:2004},
which reduces the false alarm rate at a given \ac{SNR} threshold.
Because of this, when we reduce the \ac{SNR} threshold of the
present pipeline to find the value that gives the same false alarm rate
as the phenomenological pipeline,
we effectively compensate for the lost signal power associated with
using non-spinning templates to search for spinning systems.

\section{Upper limits neglecting spin}\label{sec:ul}

In the absence of detection, we set upper limits on the rate
of \ac{CBC} per unit $L_{10}$, for several canonical binary systems
and as a function of mass of the compact binary system.

For each mass range of interest, we calculate the rate upper limit at
$90\%$ \ac{CL} using the loudest event formalism \cite{loudestGWDAW03,ul},
described in Appendix \ref{appendix:post}.
In the limit where the loudest event is consistent with the background,
the upper limit we obtain tends toward
$\mathcal{R}_{90\%} \sim 2.303 / (T {\cal C}_{L})$,
where $T$ is the total observation time (in years) 
and ${\cal C}_{L}$ is the cumulative luminosity 
(in $L_{10}$) to which this search is sensitive above its loudest event.
We derive a Bayesian posterior distribution for the rate
as described in Ref.~\cite{ul}.

In order to evaluate the cumulative luminosity
we multiply the detection efficiency, as a function of mass and distance,
by the luminosity calculated from a galaxy population \cite{LIGOS3S4Galaxies}
for the nearby universe. 
The cumulative luminosity is then this product integrated over distance.
The cumulative luminosity for this search can be found in
Table \ref{tab:ng}.


\begin{table}[t]
\center
\begin{tabular}{c | c | c | c}
\hline \hline
\multicolumn{1}{m{5cm}|}{\centering Coincidence Time} & H1H2L1 & H1L1 & H2L1 \\
\hline
\multicolumn{1}{m{5cm}|}{\centering Observation Time (yr)} & 0.336 & 0.020 & 0.041 \\
\hline
\multicolumn{1}{m{5cm}|}{\centering Cumulative Luminosity $\left({L_{10}}\right)$} & $\sim 250$ & $\sim 230$ & $\sim 120$ \\
\hline
\multicolumn{1}{m{5cm}|}{\centering Calibration Error} & 21\% & 3.9\% & 16\% \\
\hline
\multicolumn{1}{m{5cm}|}{\centering Monte Carlo Error} & 5.4\% & 16\% & 13\% \\
\hline
\multicolumn{1}{m{5cm}|}{\centering Waveform Error} & 26\% & 11\% & 20\% \\
\hline
\multicolumn{1}{m{5cm}|}{\centering Galaxy Distance Error} & 14\% & 13\% & 6.1\% \\
\hline
\multicolumn{1}{m{5cm}|}{\centering Galaxy Magnitude Error} & 17\% & 17\% & 16\% \\
\hline
\multicolumn{1}{m{5cm}|}{\centering $\Lambda$ [Eq.~\ref{eqn:lambda}]} & 0.30 & 0.41 & 0.72 \\
\hline
\hline
\end{tabular}
\caption{Detailed Results of the BNS Upper Limit Calculation}
Summary of the search for BNS systems.  The observation time is reported
after category 3 vetoes.
The cumulative luminosity is the luminosity to which the search is sensitive
above the loudest event for each coincidence time and is rounded to two
significant figures.
The errors in this table are listed as logarithmic errors in the luminosity
multiplier based on the cited sources of error.
\label{tab:BNSng}
\end{table}

\begin{table}[t]
\center
\begin{tabular}{c | c | c | c}
\hline \hline
\multicolumn{1}{m{3cm}|}{\centering System} & BNS & BBH & BHNS \\
\hline
\multicolumn{1}{m{3cm}|}{\centering Component Masses $\left(M_{\odot}\right)$} & 1.35/1.35 & 5.0/5.0 & 5.0/1.35 \\
\hline
\multicolumn{1}{m{3cm}|}{\centering $D_{\rm horizon}$ $\left({\rm Mpc}\right)$} & $\sim 30$ & $\sim 80$ & $\sim 50$ \\
\hline
\multicolumn{1}{m{3cm}|}{\centering Cumulative Luminosity $\left({L_{10}}\right)$} & 250 & 4900 & 990 \\
\hline
\multicolumn{1}{m{3cm}|}{\centering $\Lambda$ [Eq.~\ref{eqn:lambda}]} & 0.30 & 0.59 & 0.45 \\
\hline
\multicolumn{1}{m{3cm}|}{\centering Marginalized Upper Limit $\left({{\rm yr}^{-1} L_{10}^{-1}}\right)$} & \BNSul & \BBHul & \BHNSul \\
\hline
\hline
\end{tabular}
\caption{Overview of Results of the Upper Limit Calculations}
Summary of the search for BNS, BBH, and BHNS systems.
The horizon distance is the distance at which an optimally oriented and
optimally located source with the appropriate mass would produce an trigger
with an \ac{SNR} of $8$ in the 4 km detectors and averaged over the search.
The cumulative luminosity from H1H2L1 time is rounded to two significant
figures.
\label{tab:ng}
\end{table}


We apply the above upper limit calculation to three canonical binary
masses as well as calculating the upper limit as a function of mass.
Our three canonical binary masses are
\ac{BNS} $(m_1 = m_2 = (1.35\pm 0.04)~\Msun)$,
\ac{BBH} $(m_1 = m_2 = (5\pm 1)~\Msun)$, and
\ac{BHNS} $(m_1 = (5\pm 1)~\Msun,~m_2 = (1.35\pm 0.04)~\Msun)$.
We use Gaussian distributions in component mass 
centered on these masses,
with standard deviations given above following the $\pm$ symbols.

We combine the results of this search from the three different observation
times in a Bayesian manner, described in Appendix \ref{appendix:post},
and the results from previous science runs \cite{LIGOS3S4all,S3_BCVSpin}
are incorporated in a similar way.

Assuming that spin is not important in these systems,
we calculate upper limits on the rate of binary coalescences using
our injection families that neglect spin (Appendix \ref{appendix:injections}).
There are a number of uncertainties which affect the upper limit,
including systematic errors associated with detector calibration,
simulation waveforms, Monte Carlo statistics, 
and galaxy catalog distances and magnitudes \cite{systematics}.
We marginalize over these, as described in Appendix \ref{appendix:errors}
and obtain upper limits on the rate of binary coalescences of
\begin{eqnarray}
\mathcal{R}_{90\%,{\rm BNS}} = \BNSul\, \mathrm{yr}^{-1}\mathrm{L_{10}}^{-1} \\
\mathcal{R}_{90\%,{\rm BBH}} = \BBHul\, \mathrm{yr}^{-1}\mathrm{L_{10}}^{-1} \\
\mathcal{R}_{90\%,{\rm BHNS}} =  \BHNSul\, \mathrm{yr}^{-1}\mathrm{L_{10}}^{-1}
\end{eqnarray}

\begin{figure*}[t]
\includegraphics[width=0.9\textwidth]{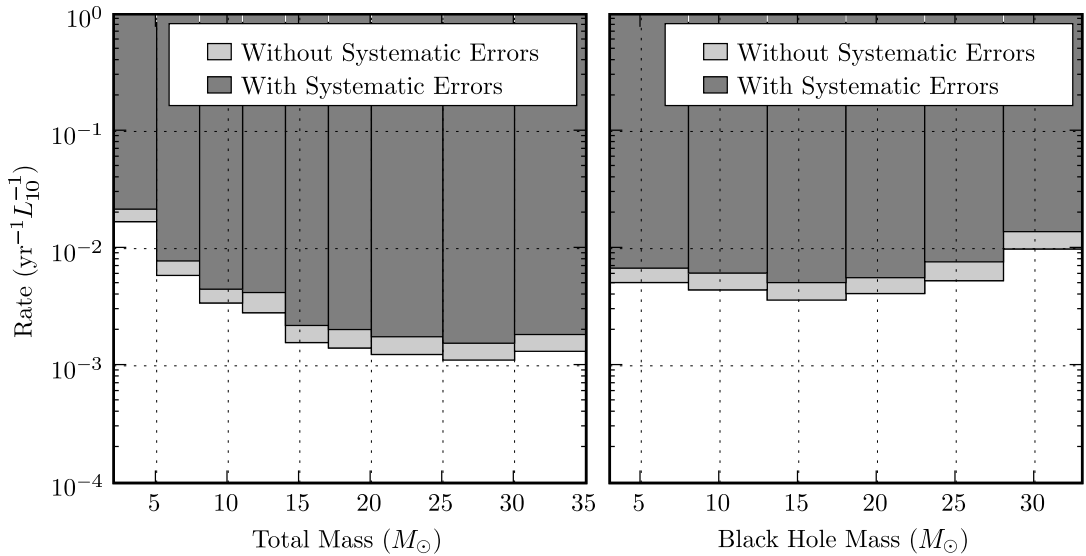}
\caption{
Upper limits on the binary coalescence rate per year and per
$L_{10}$ as a function of total mass of the binary system with a
uniform distribution in the mass ratio (left) and as a
function of the mass of a black hole in a \ac{BHNS} system with a neutron star
mass of $1.35~\Msun$ (right).
The darker area shows the excluded region after accounting for marginalization
over the estimated systematic errors.
The lighter area shows the additional region that would have been if the
systematic errors had been ignored.
}
\label{fig:upperlimit}
\end{figure*}

We also calculate upper limits for two additional cases:
as a function of the total mass of the binary,
with a uniform distribution in the mass ratio $q = m1 / m2$,
and as a function of the mass of the black
hole in a \ac{BHNS} system, holding fixed the mass of the neutron star
at $m_{\rm NS} = 1.35~\Msun$ (Fig.~\ref{fig:upperlimit}).

\section{Upper limits including spin}\label{sec:spininj}
Above we have reported upper limits on the rate of mergers for different
classes of objects using injection waveforms generated assuming non-spinning
objects.
We can also evaluate the upper limits using injection waveforms that take into
account the effects of spinning bodies.

Since the maximum possible rotational angular momentum $S$
for a black hole of mass $m$ is $Gm^2/c$,
it is useful to describe the spin of a compact object in
terms of the dimensionless spin parameter
$\hat{a} = \left({c S}\right) / \left({G m^2}\right)$.
The distribution of black hole spin magnitudes within the range
$0 \le \hat{a} \le 1$,
as well as their orientations relative to binary orbits,
is not well constrained by observations.
To illustrate the possible effects of BH spins on our sensitivity to
\ac{BBH} and \ac{BHNS} signals,
we provide an example calculation using a set of injections
of signals simulating systems whose component objects have $\hat{a}$
uniformly distributed between 0 and 1 (Appendix \ref{appendix:injections}).
On the other hand, assuming a canonical mass and uniform density,
astrophysical observations of neutron stars
show typical angular momenta corresponding to
$\hat{a}\ll 1$ \cite{ATNF:psrcat}.
In addition, the spin effects are found to be weak for the frequency range
of interest for \ac{LIGO} \cite{Apostolatos:1994}
so the \ac{BNS} upper limits in Section \ref{sec:ul}
are valid even though we have ignored the effects of spin.

Using the above injections, we obtain marginalized upper limits on the rate of
binary coalescences of
\begin{eqnarray}
\mathcal{R}_{90\%,{\rm BBH}} = \SBBHul\, \mathrm{yr}^{-1}\mathrm{L_{10}}^{-1} \\
\mathcal{R}_{90\%,{\rm BHNS}} =  \SBHNSul\, \mathrm{yr}^{-1}\mathrm{L_{10}}^{-1}
\end{eqnarray}

\section{Conclusions}\label{sec:conc}

We have searched for gravitational waves from coalescing
compact binary systems with total mass ranging from
$2$ to $35~\Msun$, using data from the first year of the
fifth science run (S5) of the three Initial LIGO detectors.
In doing so, we have investigated the efficacy of searching for \ac{BBH}
signals with 2\ac{PN} \ac{SPA} non-spinning templates and have found them to be
effective even at the relatively high total mass of $35~\Msun$.
Additionally, we have found the non-spinning templates can effectively
capture spinning signals with some loss of efficiency.
The result of the search was that no plausible gravitational wave signals
were observed above the background.
We set upper limits on the rate of these types of events
that are two orders of magnitude smaller than the previous
observational upper limits \cite{LIGOS3S4all,S3_BCVSpin},
although they are still several orders of magnitude
above the range of astrophysical estimates \cite{Kalogera:2004nt,Kalogera:2004tn,OShaughnessy:2005,Oshaughnessy:2006b}.
In the coming years, LIGO and other ground-based detectors
will undergo significant upgrades.
We expect to be able to significantly improve our sensitivity 
to gravitational waves from compact binary coalescences
and are preparing for the first detections and studies.

\acknowledgments

The authors thank Marie Anne Bizouard for her time and help 
in reviewing the accuracy of this search.
The authors gratefully acknowledge the support of the United States
National Science Foundation for the construction and operation of the
LIGO Laboratory and the Science and Technology Facilities Council of the
United Kingdom, the Max-Planck-Society, and the State of
Niedersachsen/Germany for support of the construction and operation of
the GEO600 detector. The authors also gratefully acknowledge the support
of the research by these agencies and by the Australian Research Council,
the Council of Scientific and Industrial Research of India, the Istituto
Nazionale di Fisica Nucleare of Italy, the Spanish Ministerio de
Educaci\'on y Ciencia, the Conselleria d'Economia, Hisenda i Innovaci\'o of
the Govern de les Illes Balears, the Royal Society, the Scottish Funding 
Council, the Scottish Universities Physics Alliance, The National Aeronautics 
and Space Administration, the Carnegie Trust, the Leverhulme Trust, the David
and Lucile Packard Foundation, the Research Corporation, and the Alfred
P. Sloan Foundation.

\appendix

\section{Data Quality Criteria}\label{appendix:dataqual}

When analyzing data from LIGO's detectors, it is important to know
the status of the detectors at different times.
We define data quality flags as time intervals containing known artifacts
introduced into the data by instrumental or environmental effects.  
We examine the correlation between triggers from an individual detector
and the \ac{DQ} flags, and if we find a correlation by comparing the rate
of triggers vetoed to the rate we would expect from fraction of time vetoed
(we call the \textit{dead-time}), we use them as vetoes.  
Our understanding of the coupling between the effect that prompted the
\ac{DQ} flag and the resulting triggers in the pipeline is measured
in part by the fraction of the \ac{DQ} flags that are used to veto triggers
(called the \textit{use percentage}). 
We define four different categories of \ac{DQ} vetoes based on the
above criteria.

We categorize \ac{DQ} vetoes as \textit{category 1 vetoes}
when we know of a severe problem with the data,
bringing in to question whether the detector was actually in observation mode.
An example case for H2 is when loud vibrations were caused in the detector
environment in order to test the response of the seismic isolation systems.
We categorize \ac{DQ} vetoes as \textit{category 2 vetoes}
when there is a known coupling between the \ac{GW} channel
and the auxiliary channel, the veto is correlated with triggers from the 
individual detector, particularly at high \ac{SNR}, and when there is a use 
percentage of 50\% or greater.
An example is when any of the data channels in the length sensing and control
servo reach their digital limit.
We categorize \ac{DQ} vetoes as \textit{category 3 vetoes}
when the coupling between the auxiliary channel and the \ac{GW} channel
is less established or when the use percentage is low, but we still
find a strong correlation between the vetoes and the triggers.
An example is when the winds at the detector site are over 30 Mph.
We categorize \ac{DQ} vetoes as \textit{category 4 vetoes}
when the coupling between the auxiliary channel and the \ac{GW} channel
is not well established, when the use percentage is low, when the overall
dead-time is several percent or greater, or the correlation is weak.
An example is when nearby aircraft pass overhead.
We compare all of these vetoes with the times of hardware injections, which 
measure the response of the detector to a simulated gravitational wave signal,
in order to confirm that the \ac{DQ} vetoes are not sensitive to real signals 
in the data. 

\begin{figure*}[t]
\includegraphics[width=0.5\textwidth]{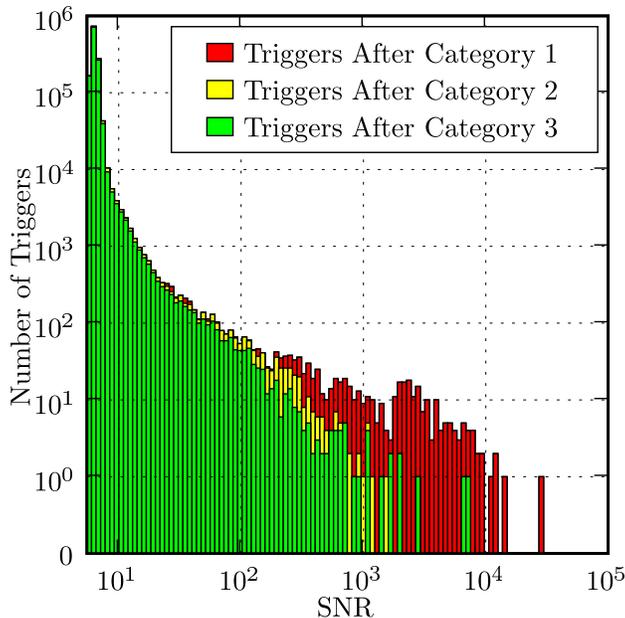}
\caption{Histogram of triggers for the H2 detector,
clustered by the trigger with maximum \ac{SNR} within 10 seconds,
plotted after removing triggers occurring during times vetoed by
category 1, 2, and 3 vetoes.
The tail of the \ac{SNR} distribution is significantly reduced
by both the category 2 and 3 vetoes.
}
\label{fig:vetoedclusters}
\end{figure*}

We do not analyze data vetoed by category 1 \ac{DQ} vetoes.
We remove triggers in times defined by category 2 and 3 \ac{DQ} vetoes 
from the upper limit calculation.
These veto categories significantly reduce the \ac{SNR} of outlying 
triggers (Fig.~\ref{fig:vetoedclusters}).
As an exercise, we follow up the loudest coincident triggers after
each category of veto is applied,
including after Category 4 vetoes
(See Appendix \ref{appendix:detection_checklist}).
This allows us to investigate the action of the vetoes
by a ``case study" method.

\section{Follow-up Procedure for Coincident Triggers}
\label{appendix:detection_checklist}

As an exercise, we check our loudest coincident triggers with a
list of tests designed to see if a statistically significant trigger
is believable as a detection candidate.
Ref.~\cite{detection-checklist-GWDAW07} describes the tests that we
perform on the trigger and the data surrounding it.
At present, our standard tests include the following:
We check the integrity of the data for corruption.
We also check the status of the detectors and the presence of any
data quality flags in the surrounding data.
We assess whether there could have been environmental or
instrumental causes found in auxiliary channels at the time of the trigger.
We check the appearance of the data at the time of the trigger in the form of
\ac{SNR} time series, the $\chi^2$ time series,
and time-frequency spectrograms.

In addition, for any statistically significant candidate that
survives the tests listed above, we plan to do the following:
Assess the coherence between the signals recorded by each
individual detector operating at the time of the event.
Verify the robustness of the trigger against small changes in the pipeline
(i.e.~changes in the adjacent Fourier transform boundaries or
changes in the calibration of the data).
Check the robustness across pipelines by employing other search techniques
to analyze the same data (i.e.~\ac{CBC} pipelines using different
templates or algorithms designed to search for unmodeled bursts).
Finally, we will check for coincidence with external searches for
gamma-ray bursts, optical transients, or neutrino events.
(This last test is for information only, as a genuine \ac{GW} event might or
might not be accompanied by other signals.)

As mentioned in Appendix \ref{appendix:dataqual},
we examine the distribution of triggers after each category of veto is applied.
In case there is a statistically significant outlier
after only Category 1 or after Categories 1 and 2 are applied,
we carry out a follow up exercise to see if the veto that eventually
rejected the event was rightfully applied.
There are two reasons that this could be important.
Firstly, a very strong gravitational wave from within the Milky Way could cause
an instrumental saturation of the sort that we use as a veto;
this kind of problem would be easy to diagnose if it were to occur,
since the signal would be strong enough for us to see in the moments
leading up to the signal-induced saturation.
Secondly, we want to guard against false dismissal of a candidate by other
kinds of vetoes, which can have non-negligible deadtime associated with them.
Some of our vetoes are associated with recognizable forms of false signals;
we check to be sure that a vetoed loud event looks like that kind of
false signal, and not like a genuine coalescence signal.
In the search described in this paper,
there was a single statistically significant outlier in the
distribution of events after the application of veto Categories 1 and 2.
The follow up exercise confirmed that the Category 3 test that vetoed
that event was correctly applied.

Before unblinding the data we discovered an error in the computation of the
template metric.  This metric is used in the placement of the bank and the
coincidence test.  The error caused the metric distance between templates to be
overestimated for the higher mass signals.  This has the effect of causing the
template placement algorithm to overcover the higher mass region (i.e.~to
produce a bank with less than the requested 3\% loss in signal-to-noise ratio).
This increased the computational cost of the search, but did not significantly
reduce the sensitivity.  However, this error also affected the coincidence
algorithm by overestimating the distance between triggers for high mass
signals.  Since the coincidence window was empirically tuned on software
injections and time-shifted coincidences the impact on the sensitivity of the
search was not significant.  Consequently, the decision was taken to unblind
the data using the original, sub-optimal analysis in order to begin studying
any possible detection candidates and to use this result to compute the upper
limit (in the absence of a detection).  The decision was also taken to perform
a complete re-analysis of the data with the corrected metric to verify the
(non-)detection statement from the original search.  The results of the
re-analysis were consistent with the original analysis and did not produce any
plausible gravitational-wave signals.

\section{Effective \ac{SNR}}\label{appendix:effsnr}

For this search we employ the same definition of combined effective \ac{SNR}
as was used in the \ac{BNS} searches of Ref.~\cite{LIGOS3S4all}.
The combined effective \ac{SNR} is constructed as follows.

The single-detector \ac{SNR} is produced by matched filtering the data
against our templates.
The complex output from the matched filter, $z$, is given by
\begin{equation}
\label{eqn:matchedfilter}
z = 4 \int\limits_0^\infty{\frac{\tilde{s}\left({f}\right)^{\ast} \tilde{h}\left({f}\right)}{S_{n}\left({f}\right)}}df,
\end{equation}
where $\tilde{s}\left({f}\right)^{\ast}$ is the complex conjugate of the
Fourier transform of the data,
$\tilde{h}\left({f}\right)$ is the Fourier transform of the template,
and $S_{n}\left({f}\right)$ is the power spectral density of the noise
in the detector.
The template normalization $\sigma$ is given by
\begin{equation}
\label{eqn:sigma}
\sigma^2 = 4 \int\limits_0^\infty{\frac{\tilde{h}\left({f}\right)^{\ast} \tilde{h}\left({f}\right)}{S_{n}\left({f}\right)}}df.
\end{equation}
The $z$ and $\sigma$ are combined to give the
single-detector \ac{SNR}, $\rho$, using
\begin{equation}
\label{eqn:snr}
\rho = \frac{|z|}{\sigma}.
\end{equation}

From $\rho$ we define the
\textit{effective \ac{SNR}},
$\rho_{\rm eff}$, as:
\begin{equation}
\label{eqn:effsnr}
\rho_{\rm eff}^2 =
\frac{\rho^2}{\sqrt{\left(\frac{\chi^2}{2p-2 }\right)\left(1+
\frac{\rho^2}{250}\right)}}, 
\end{equation}
where $p$ is the number of bins used in the $\chi^2$ test,
which is a measure of how much the signal in the data looks like the
template we are searching for.
In the effective \ac{SNR}, we normalize the $\chi^2$ by $2p - 2$
since it is the number of degrees of freedom of this test.

We then combine the effective \ac{SNR}s for the single-detector triggers
that form a coincident trigger into the \textit{combined effective \ac{SNR}},
$\rho_{\rm c}$, for that coincident trigger using:
\begin{equation}
\label{eq:combinedstat}
\rho_{\rm c}^2=\sum_{i=1}^N \rho_{{\rm eff}, i}^2.
\end{equation}

This definition of the combined effective \ac{SNR} reduces the apparent
significance of non-Gaussian instrumental artifacts since
it weights the \ac{SNR} by the $\chi^2$.
This effectively cuts down on outliers from the expected \ac{SNR}
distribution due to Gaussian noise.
In addition, we test this definition of the combined effective \ac{SNR} using
software injections and find it does not significantly affecting the apparent
significance of real signals.

\section{False Alarm Rate}\label{appendix:FAR}

Previously \cite{LIGOS3S4all}, we defined the loudest event for the entire
parameter space based on the combined effective \ac{SNR},
$\rho_{\rm c}$ (Appendix \ref{appendix:effsnr}).
Since we are searching over a larger portion of parameter space than
before, we find that the distribution of combined effective \ac{SNR} for
time-shifted coincident triggers
varies significantly over different portions of the parameter space.
In general, this seems to be affected by two factors.
We see a suppression of the combined effective \ac{SNR} distributions
for time-shifted coincident triggers when looking at triple coincident
triggers compared to double coincident triggers.
Also, we find smaller combined effective \ac{SNR} distributions for
time-shifted coincident triggers in the lower mass regions than in the
higher mass regions.

For this search, we have decided to divide the parameter space
into regions with similar combined effective \ac{SNR} distributions for
time-shifted coincident triggers.
We separate the triggers into different categories,
where the categories are defined by the mean template masses of the triggers
and triggers types
(triple coincident triggers found in triple coincident time,
double coincident triggers found in triple coincident time,
and double coincident triggers found in double coincident time).
The categories for this search are given by the combination of
three template mass regions with divisions in chirp mass at
$M_{\rm c} = (3.48,~7.40)~\Msun$
with trigger types given by H1H2L1, H1L1, and H2L1 triggers from H1H2L1
triple coincident time,
H1L1 triggers from H1L1 double coincident time,
and H2L1 triggers from H2L1 double coincident time.

Within each category, the time-shifted coincident triggers provide an estimate
of the \ac{FAR} for each in-time coincident trigger.
When we recombine the categories from the same observation time,
the \ac{FAR} of each trigger
then needs to be normalized by the number of trials
(i.e.~the number of categories).
This normalization bestows a \ac{FAR} of $1 / T$ with the meaning
that during the observation time covered by this
search ($T$), there is expected to be a single coincidence trigger
due to background with a combined effective \ac{SNR} at that level.

The \ac{IFAR} is used as our detection statistic and
in-time coincident triggers with the largest \ac{IFAR} 
(across all categories) are our best detection candidates.

\section{Simulated Waveform Injections}\label{appendix:injections}

In order to measure the efficiency of our pipeline to recovering
\ac{GW} signals from \ac{CBC}, we inject several different \ac{PN}
families of waveforms into the data and check to see the fraction of signals
that are recovered.
The different waveform families used for injections in this search include
GeneratePPN computed to Newtonian order in amplitude and
2\ac{PN} order in phase using formulae from Ref.~\cite{Blanchet:1996pi},
EOB computed to Newtonian order in amplitude and
3\ac{PN} order in phase using formulae from Refs.~\cite{BuonannoDamour:1999, BuonannoDamour:2000, DamourJaranowskiSchaefer:2000, Damour:2001},
Pad\'{e}T1 computed to Newtonian order in amplitude and
3.5\ac{PN} order in phase using formulae from Refs.~\cite{Damour:2000zb, Damour:1998zb},
and
SpinTaylor computed to Newtonian order in amplitude and
3.5\ac{PN} order in phase using formulae from Refs.~\cite{BuonannoChenVallisneri:2003b}
and based upon Refs.~\cite{kidder:821, Blanchet:1995ez, Blanchet:1996pi, PhysRevD.54.1417, Blanchet:2001ax, PhysRevD.47.R4183, Damour:2000zb, Apostolatos:1994};
using code from Ref.~\cite{LAL}.
Each of these families except for SpinTaylor ignores the effects of spin
on the orbital evolution.

Each of these waveform families are injected from a distribution
uniform in sky location (right ascension, declination),
uniform in the cosine of the inclination angle ($\iota$),
and uniform in polarization azimuthal angle ($\psi$).
Each of these waveform families are injected from a distribution uniform
in the total mass of the system.
Each of these waveform families are also injected uniform in $\log_{10}D$
where $D$ is the physical distance from the Earth to the source in Mpc.
This non-physical distance distribution was chosen in order to test our
pipeline on a large range of signal amplitudes.

For the SpinTaylor waveform family, each of the component objects' spin
magnitudes are chosen from a distribution uniform in the
unitless spin parameter
$\hat{a} \equiv \left({c S}\right) / \left({G m^2}\right)$,
ranging from 0 to 1.
The component objects' spin orientations relative to the initial orbital
angular momentum are chosen from a distribution uniform on a sphere.

\section{Chirp Distance}\label{appendix:chirp_distance}

In the adiabatic regime of binary inspiral, 
gravitational wave radiation is modeled accurately. 
We make use of a variety of approximation techniques
\cite{Blanchet:1995ez,Blanchet:1995,Blanchet:1996pi,Blanchet:2001ax,
Blanchet:2004ek,Damour:1998zb,BuonannoDamour:1999,BuonannoDamour:2000,
Damour:2000zb} which rely, to some extent, on the slow motion of the compact
objects which make up the binary. 
We can represent the known waveform by:
\begin{equation}\label{eq:h}
h(t) = \frac{1 {\rm Mpc}}{D_{\rm eff}} A(t)\cos{\left(\phi(t) - \phi_0 \right)}
\end{equation}
where $\phi_0$ is some unknown phase. For this search the functions $A(t)$ and
$\phi(t)$ are the Newtonian amplitude and 2\ac{PN} phase evolution respectively,
which depend on the masses and spins of the binary.

The template matched filtering will identify the masses and
coalescence time of the binary but not its physical distance $D$.
The signal amplitude received by the detector depends on the detector
response functions $F_+$ and $F_\times$,
and the inclination angle of the source $\iota$, which are unknown.
We can only obtain the \textit{effective distance} $D_{\rm eff}$,
which appears in Eq.~(\ref{eq:h}) defined as \cite{thorne.k:1987}:
\begin{equation}\label{eq:Deff}
D_{\rm eff} = \frac{D}{\sqrt{F_+^2 (1 + \cos^2\iota)^2/4 + F_\times^2
( \cos\iota)^2}} \; .
\end{equation}
The effective distance of a binary may be larger than its physical
distance.

Since the amplitude of a gravitational wave scales as the chirp mass
$M_{\rm c}$ to the five sixths power,
it is convenient to normalize the effective distance by this,
obtaining the \textit{chirp distance}, which is given by:
\begin{equation}
D_{\rm chirp} = D_{\rm eff} \left( {\frac{M_{\rm c,BNS}}{M_{\rm c}}} \right)^{\frac{5}{6}}.
\end{equation}
where $M_{\rm c,BNS}$ is the chirp mass of a canonical \ac{BNS} system.
This distance is useful in evaluating the efficiency of the search as a
function of distance since the efficiency will then be
approximately independent of mass.

\section{Posterior and Upper Limit Calculation}\label{appendix:post}

Calculating an upper limit on a rate of coalescences in the loudest-event
formalism requires knowledge of the cumulative luminosity to which the
search is sensitive and a measure of the likelihood that the loudest event
was due to the observed background.
The cumulative luminosity quantifies the potental sources of observable
\ac{CBC}, as measured by blue-light luminosity of the galaxies, which can be
detected by our search.
It is calculated by multiplying the efficiency of signal recovery for the
search as a function of distance by the physical luminosity as a function of
distance and integrating their product over distance.
We combine these with the time analyzed to calculate the posterior
on the rate for the search.
This is, assuming a uniform prior on the rate, given by \cite{ul}:
\begin{equation}
\label{eqn:posterior}
p \left({ \mu | {\cal C}_{L}, T, \Lambda }\right) = \frac{{\cal C}_{L} T}{1 + \Lambda} \left({1 + \mu {\cal C}_{L} T \Lambda}\right) e^{-\mu {\cal C}_{L} T}
\end{equation}
where $\mu$ is the rate, ${\cal C}_{L}$ is the cumulative luminosity,
$T$ is the analyzed time, and $\Lambda$ is a measure of the likelihood
of detecting a single event with loudness parameter $x$ versus such an event
occurring due to the experimental background, given by \cite{ul}:
\begin{equation}
\Lambda \left({ x }\right) = \left({ \frac{-1}{{\cal C}_{L}} \frac{d {\cal C}_{L}}{dx} }\right) \left({ \frac{1}{P_0} \frac{d P_0}{dx} }\right)^{-1}
\label{eqn:lambda}
\end{equation}

The posterior (\ref{eqn:posterior}) assumes a known value of ${\cal C}_{L}$
associated with the search.
In reality, ${\cal C}_{L}$ has associated with it systematic uncertainties,
which we model as unknown multiplicative factors,
each log-normally distributed about $1$ with errors described in
Appendix \ref{appendix:errors}.
The widths of those distributions are given in Table \ref{tab:BNSng}.
Marginalizing over all of those unknown factors,
and thus over ${\cal C}_{L}$,
gives a marginalized posterior:
\begin{equation}
p \left({ \mu | T, \Lambda}\right) = \int{ p_{d} \left({{\cal C}_{L}}\right) p \left({ \mu | {\cal C}_{L}, T, \Lambda }\right)} d{\cal C}_{L}
\end{equation}
where $p_{d} \left({{\cal C}_{L}}\right)$ is the combined probability
distribution function for ${\cal C}_{L}$ given all of those unknown factors.

The results of several experiments (e.g.~different types of S5
observing time and previous runs such as S3 and S4)
can be combined by taking the product of their likelihood functions;
in the case of uniform priors,
this is equivalent to taking the product of their posteriors,
allowing us to define the rate upper limit
$\mu$ at a confidence level $\alpha$ by solving:
\begin{equation}
\alpha = \int_{0}^{\mu}{\prod_{i}{p_{i} \left({\mu'}\right)}}d\mu'
\end{equation}
where the $p_{i} \left({\mu'}\right)$ are the marginalized posteriors from
different experiments calculated using a uniform prior on the rate.

\section{Systematic Error Calculation}\label{appendix:errors}

Systematic errors associated with \ac{CBC} searches for \ac{GW} signals include
errors associated with detector calibrations,
simulation waveforms, Monte Carlo statistics, 
and galaxy catalog distances and magnitudes.
Calculating these errors in terms of the cumulative luminosity
is described below \cite{systematics}.

We refer to statistical errors associated with the efficiency calculation as
\textit{Monte Carlo errors}.
Since we calculate the efficiency as a function of distance,
we calculate the error for a particular distance bin using the binomial formula,
which gives an error of zero when the efficiency is zero or one,
or when there are no injections in that bin.
This error is then multiplied by the physical luminosity as a function of
distance and integrated over distance to get the Monte Carlo error
in units of luminosity.

\textit{Calibration errors} in the detectors are errors on the amplitude
of the noise floor.
These errors affect the amplitude, and in turn the distance,
at which we made injections to calculate the efficiency of our search,
since the injections we made assuming a specific value of the noise floor.
The one-sigma uncertainty in the amplitude (and thus the distance)
associated with the calibration was
$8.1\%$ for H1, $7.2\%$ for H2, and $6.0\%$ for L1.
We use these numbers to calculate the calibration error given in
Table \ref{tab:BNSng} in units of luminosity by combining the logarithmic errors
in quadrature.

\textit{Waveform errors} are associated with how different the true signals
are from what we use to measure the efficiency of our pipeline
(i.e.~the mismatch between the true signals and our injections).
This error effectively reduces the distances in our efficiency calculation
since we don't recover all of the power available in the signal due to the
mismatch between the signal and our injections.
We calculate the waveform error in units of luminosity assuming
a waveform mismatch of $10\%$\cite{Damour:2000zb, pan:024014}.

\textit{Galaxy errors} are errors associated with our
galaxy catalog \cite{LIGOS3S4Galaxies}
used to construct the physical luminosity.
Galaxy errors come in two types: \textit{distance errors} and
\textit{magnitude errors}.
To calculate the error on the luminosity due to distance errors,
the physical luminosity calculation is modified such that the
galaxies' distances are increased by a factor $1 + \kappa_{j}$,
where $\kappa_{j}$ is the uncertainty in the $j^{\rm th}$ galaxy's distance
given in the galaxy catalog.
The galaxies' luminosities are also increased by a factor
$\left({1 + \kappa_{j}}\right)^2$ since the galaxies' luminosity
is only known in terms of its magnitude and distance.
To calculate the error on the luminosity due to magnitude errors,
the physical luminosity calculation is changed such that the galaxies'
luminosities are increased by an amount associated with the 
magnitude errors given in the galaxy catalog.

\section{Spinning Search Comparison}\label{appendix:BCVSpin}

The \ac{SPA} template waveforms used in this search
and described in this paper do not take spin into account. 
A phenomenological template family to search for spinning black hole
and neutron star binaries was developed in \cite{BuonannoChenVallisneri:2003b},
referred to as \textit{BCVSpin},
and has been used in a search of S3 data \cite{S3_BCVSpin}.
Using both of these template banks to compute the efficiency of recovering
signals from spinning waveforms for the different search methods,
we find that for a comparable number of false alarms,
\ac{SPA} and BCVSpin have approximately the same efficiencies,
implying it is not necessary to perform a search using BCVSpin templates
in order to target spinning signals.
The comparison of searches for spinning binaries using different signal models
and template banks is discussed further in Ref.~\cite{VDBroeck:2008}.

\begin{figure*}[t]
\includegraphics[width=0.9\textwidth]{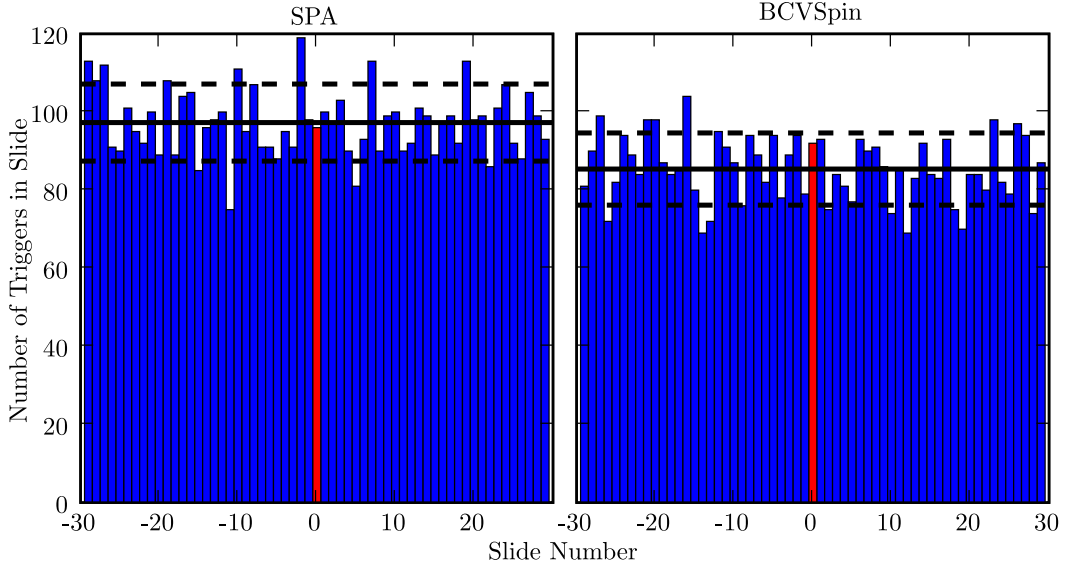}
\caption{Histograms of the number of time-shifted coincident triggers for
\ac{SPA} (left) and BCVSpin (right) templates,
in about 9 days of playground data from H1 and L1.
In this investigation,
we applied an \ac{SNR} threshold of 5.5 for \ac{SPA} and 8 for BCVSpin,
the number of triggers is approximately the same in the two cases.
}
\label{fig:backgroundrates}
\end{figure*}

\begin{figure*}[t]
\includegraphics[width=0.9\textwidth]{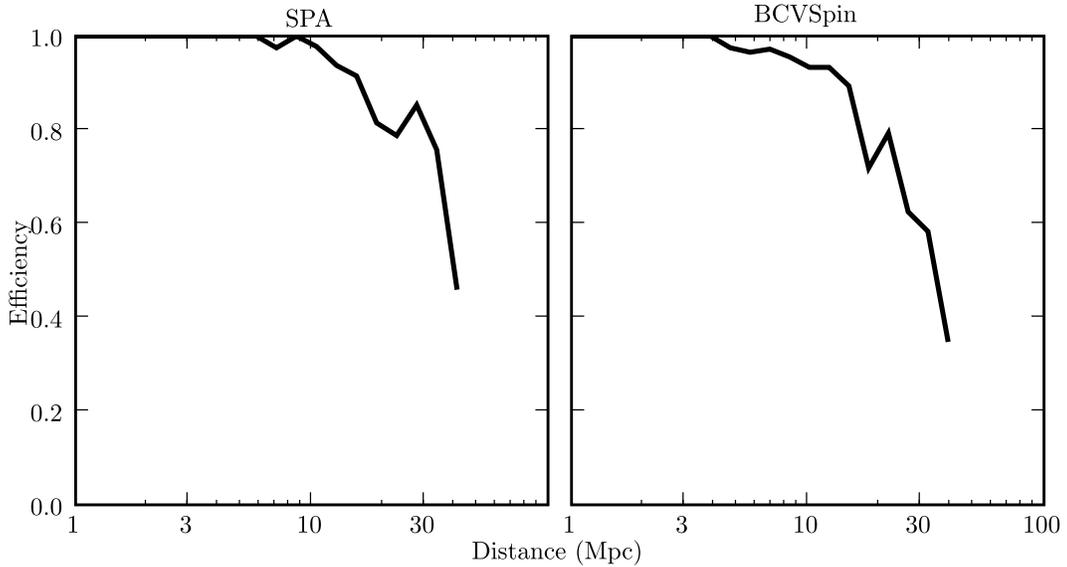}
\caption{Injection recovery efficiencies plotted against distance for \ac{SPA}
(left) and BCVSpin (right) templates.
}
\label{fig:efficiencies}
\end{figure*}

What is important for a search is how efficient
banks are in picking up signals in the data. Given a large number
of injections in the data, the efficiency is the ratio of the
number of found injections to the total number of injections made.
A fair comparison requires that efficiencies be evaluated
\emph{for the same \ac{FAR}}.
To estimate the background rate, we counted the number of coincident
triggers in time-shifted data between H1 and L1.

The \ac{SNR} for BCVSpin involves six degrees of freedom,
compared to only two for \ac{SPA}.
As a consequence, BCVSpin picks up glitches more easily,
and to have the same background rate as for \ac{SPA}, it needs a higher
\ac{SNR} threshold. (This problem had already been pointed out and discussed
in \cite{BuonannoChenVallisneri:2003b}; here we are seeing it in real data.)
It was found that \ac{SPA} with an \ac{SNR} threshold
of 5.5 and BCVSpin with an \ac{SNR} threshold of 8 lead to comparable
\ac{FAR}s (Fig.~\ref{fig:backgroundrates}).  

With these \ac{SNR} thresholds, we are in a position to compare the
efficiencies of \ac{SPA} and BCVSpin banks for a given \ac{FAR}.
For our purposes,
an injection will be considered found if it had an \ac{SNR} above the
chosen threshold with at least one template in the bank, within a
certain time interval around the time when the injection was actually made.
In the case of \ac{SPA}, the width of this interval can be chosen to be 40 ms.
BCVSpin templates, being phenomenological, turn out to have a larger
timing inaccuracy, and an interval of 100 ms was found to be more appropriate.
We made 1128 injections distributed logarithmically in distance
between 1 Mpc and 50 Mpc, with component masses randomly chosen
between $1\,M_\odot$ and $30\,M_\odot$ but restricting total mass to
$30\,M_\odot$, component spin magnitudes
$0.7 < \hat{a}_{i} < 1$, $i=1,2$,
and arbitrary directions for the initial spin vectors.
For the \ac{SNR} thresholds of 5.5 for \ac{SPA} and 8 for BCVSpin,
in H1 the efficiency of \ac{SPA} came out to be 0.93, versus 0.89
for BCVSpin; for L1 the numbers are similar.
Fig.~\ref{fig:efficiencies} shows the efficiencies binned in distance.
Hence, for comparable \ac{FAR}s, \ac{SPA} and BCVSpin have
approximately the same efficiencies showing it is not necessary
to perform a search using BCVSpin templates in order
to target spinning signals.

\bibliography{../bibtex/iulpapers}

\end{document}